\documentclass{iau_FM}
\usepackage{graphicx,color}

\title[FM 8.~Magnetic fields, ISM structure and galactic outflows]
{Magnetic field effects on the ISM structure and galactic outflows}

\author[Shukurov \etal]   
{A.~Shukurov$^1$, C.~C.~Evirgen$^1$, A.~Fletcher$^1$, P.~J.~Bushby$^1$
 \and F.~A.~Gent$^2$}

\affiliation{$^1$School of Mathematics, Statistics and Physics, Newcastle University,
Newcastle upon Tyne, NE1~7RU, U.K.\\ 
emails: {\tt anvar.shukurov@ncl.ac.uk (AS), c.c.evirgen@ncl.ac.uk (CCE), 
andrew.fletcher@ncl.ac.uk (AF), paul.bushby@ncl.ac.uk (PJB)} \\[\affilskip]
$^2$Department of Computer Science, Aalto University, Oteniemi, ESPOO, Finland \\
email: {\tt frederick.gent@aalto.fi}}

\pubyear{2018}
\jname{Astronomy in Focus, Volume 1} 
\editors{L.~Feretti, G.~Heald, M.~Johnston-Hollitt \& F.~Govoni, eds}
\begin{document}

\maketitle
\begin{abstract}
The role of magnetic fields in the multi-phase interstellar medium (ISM) is explored
using magnetohydrodynamic (MHD) simulations that include energy injection by supernova
(SN) explosions and allow for dynamo action. Apart from providing additional pressure
support of the gas layer, magnetic fields reduce the density contrast between the warm 
and hot gas phases and quench galactic outflows. A dynamo-generated, self-consistent
large-scale magnetic field affects the ISM differently from an artificially imposed, 
unidirectional magnetic field.
\end{abstract}
Our understanding of global magnetic effects in spiral galaxies remains limited despite 
a growing appreciation of their significance. Of particular importance is the role of
magnetic fields (and of the cosmic rays controlled by them) in galactic outflows and the
resulting feedback of star formation on galactic evolution. Studies of magnetic effects in 
the ISM are hindered by its complex, multi-phase structure and by the complexity of the
conversions between kinetic and magnetic energies in random flows (addressed by dynamo 
theory). Furthermore, observational estimates of magnetic field strength are restricted to 
the warm and cold ISM phases whereas galactic outflows are driven by the hot gas.

Simulations of the SN-driven ISM have reached a sufficient level of physical adequacy
to use them as numerical experiments to explore various effects inaccessible to direct
observations. Among such simulations, only a few contain the ingredients (especially
differential rotation and stratification) required to simulate large-scale dynamo action
(\cite[Gressel \etal\ 2008]{GEZR08}, \cite[Gent \etal\ 2013]{GSFSM13}).  

We analyse a numerical model of the local ISM, developed by 
\cite[Gent \etal\ (2013)]{GSFSM13}, that produces an exponentially growing magnetic field 
at a scale comparable to the size of the computational domain (1\,kpc), starting from
a dynamically negligible seed magnetic field and reaching a statistically steady state
where magnetic, random kinetic and thermal energy densities are comparable in magnitude.
The ISM is represented in three phases separated in terms of specific entropy
(or density and temperature) while magnetic field consists of both the mean and 
fluctuating parts. The SN activity supports a systematic outflow away from the mid-plane. 
Comparing the ISM properties in the initial and late stages, where magnetic fields are, 
respectively, negligible and dynamically significant, we can assess the role of magnetic 
fields in the structure and properties of the ISM. Both stages represent statistically 
steady states without and with magnetic field, respectively. The most significant magnetic
effects are as follows \cite[(Evirgen \etal\ 2018)]{EGSFB18}.

\begin{table}
 \begin{center}
 \caption{ Fits to the outflow speed, of the form (1), at  various distances $|z|$ from 
 the mid-plane.}
 \label{V_z}
 {\scriptsize
    \begin{tabular}{lccc}
\hline
Distance to the mid-plane [kpc]         &$V_0$ [km\,s$^{-1}$]  &$\xi$ &$n$\\
\hline
\phantom{$0.15<$} $|z|<0.15$ 		    & 11                    & 4.4   & 2.7 \\
$0.15<|z|<0.3$	                        & 17                    & 2.4   & 2.4 \\
\phantom{0}$0.3<|z|<0.6$	            & 17                    & 3.9   & 1.9 \\
\phantom{0}$0.6<|z|<1.0$	            & 11                    & 2.5   & 1.5 \\
\hline
    \end{tabular}
  }
 \end{center}
\end{table}

{\underline{\it The multi-phase structure}}. In agreement with the topological analysis of
\cite[Makarenko \etal\ (2018)]{MSHRBF18}, we show that strong local magnetic fields 
efficiently confine SN remnants, leading to a lower fractional volume of the hot gas 
(decreasing from 25\% to 9\% near the mid-plane as magnetic field grows), its lower 
temperature and higher density (by a factor of 3--10), whereas the total mass of the hot 
gas varies little. As a result, the density contrast between the warm and hot phases is
reduced by almost an order of magnitude.

{\underline{\it Outflow speed}}. As the magnetic field grows, the mean speed of the 
systematic gas outflow $V_z$ away from the mid-plane decreases from about 20\,km\,s$^{-1}$ 
for a weak magnetic field to just a few km\,s$^{-1}$ when magnetic field becomes strong.
The decrease is rather abrupt: $V_z$ is only weakly dependent upon the magnetic field when 
the mean magnetic field $B$ is weaker than that corresponding to energy equipartition with 
the random gas flow, $B_{\rm eq}=(4\pi\rho v^2)^{1/2}$, where $\rho$ is the gas mass 
density and $v$ the root-mean-square random speed. However, it then decreases as a rather 
high power of $B$,\vspace*{-5pt}
$$\vspace*{-5pt}
V_z\simeq V_0\left[1+\xi(B/B_{\rm eq})^n\right]^{-1}\,,\eqno{(1)}
$$
with the values of $V_0$, $\xi$ and $n$ given in Table~\ref{V_z}. In similar simulations,
\cite[Bendre \etal\ (2015)]{BGE15} also find that the outflow is suppressed by the
large-scale magnetic field although they find a weaker dependence of $V_z$ on $B$
with $n=2$ (and $V_0\approx13\,{\rm km\,s^{-1}}$ and $\xi\approx1.5$ at 
$|z|=0.8\,{\rm kpc}$). The value of $V_0$ depends on the supernova rate $\nu$; 
\cite[Bendre \etal\ (2015)]{BGE15} find $V_0\propto \nu^{0.4}$.
We note that the value of $V_0$ at large $|z|$ is likely to be sensitive to the size of the
computational domain: simulations of \cite[Bendre \etal\ (2015)]{BGE15} in $|z|\leq2\,$kpc 
find larger values of $V_z$ at $|z|\simeq1\,$kpc than in our model where $|z|\leq1\,$kpc.

{\underline{\it The importance of dynamo action}}. The sensitivity of the ISM structure
and outflow speed to magnetic field is only obtained in simulations that admit large-scale
dynamo action (\cite[Bendre \etal\ 2015]{BGE15}, \cite[Evirgen \etal\ 2018]{EGSFB18}).
Broadly similar simulations where dynamo action is precluded and a unidirectional
magnetic field is imposed (e.g., \cite[de Avillez \& Breitschwerdt  2005]{dAB05},
\cite[Henley \etal\ 2015]{HSKHML15}, \cite[Girichidis \etal\ 2016]{GWNGWGKCPDB16}) do 
not reveal any similar magnetic effects. This contradiction requires careful analysis;
it suggests rather strongly that dynamo action that produces a large-scale magnetic field
responding \textit{self-consistently} to the ISM environment is essential to capture reliably
magnetic effects on the ISM.

\vspace*{-6pt}


\end{document}